\documentclass[aps,prl,twocolumn,groupedaddress]{revtex4}
\usepackage{graphicx}
\usepackage{color}
\usepackage{amsmath}
\usepackage{amsfonts}
\usepackage{bm}
\newcommand{\ket}[1]{\ensuremath{\left|{#1}\right\rangle}}
\newcommand{\bra}[1]{\ensuremath{\left\langle{#1}\right |}}

\begin{document}


\title{Detection of transverse entanglement in phase space}

\author{D. S. Tasca}
\email[]{tasca@if.ufrj.br}
\affiliation{Instituto de F\'{\i}sica,
Universidade Federal do Rio de Janeiro, Caixa Postal 68528, Rio de
Janeiro, RJ 21941-972, Brazil}
\author{S. P. Walborn}
\affiliation{Instituto de F\'{\i}sica, Universidade Federal do Rio
de Janeiro, Caixa Postal 68528, Rio de Janeiro, RJ 21941-972,
Brazil}
\author{F. Toscano}
\affiliation{ Funda\c{c}\~ao Centro de Ci\^encias e
              Educa\c{c}\~ao Superior a Dist\^ancia do Estado
              do Rio de Janeiro,
              20943-001 Rio de Janeiro, Brazil}
 \affiliation{Instituto de F\'{\i}sica, Universidade Federal do Rio
              de Janeiro, Caixa Postal 68528, Rio de Janeiro, RJ 21941-972,
              Brazil}
\author{P. H. Souto Ribeiro}
\affiliation{Instituto de F\'{\i}sica,
Universidade Federal do Rio de Janeiro, Caixa Postal 68528, Rio de
Janeiro, RJ 21941-972, Brazil}
\date{\today}

\begin{abstract}
Transverse entanglement between pairs of photons can be detected
through intensity correlation measurements in the near and far
fields. We show theoretically and experimentally that at
intermediate zones, it is also possible to detect transverse
entanglement performing only intensity correlation measurements.
Our results are applicable to a number of physical systems.
\end{abstract}


\maketitle


Detection and quantification of entanglement is essential for the
development of many applications in the field of quantum
information. Several tasks proposed to take
advantage of the entanglement properties of quantum systems can be
experimentally tested with photons produced from
spontaneous parametric down-conversion(SPDC) \cite{bouwmeester00}.
This is a versatile system, since SPDC photons can be prepared
in entangled states of many different degrees of freedom, such as
polarization \cite{kwiat95}, time-bins \cite{tapster94},
orbital angular momentum \cite{vaziri02}, as well as
transverse spatial variables \cite{howell04,dangelo04}.
The latter concerns correlations between the transverse components of
the wave vectors of the signal and idler photons, which have been extensively studied and utilized in the last decade
\cite{ribeiro94b,strekalov95,abouraddy01}. They arise due to the localization of the emission of
photon pairs and the phase matching conditions for the non-linear
interaction between the pump, signal and idler fields. Even though
the quantum nature of spatial correlations was already evident
\cite{nogueira01}, the formal relationship with
entanglement has only been demonstrated a few years ago
\cite{howell04,dangelo04}. Transverse entanglement was detected
through the violation of a non-separability criteria
\cite{duan00,mancini02}, based on intensity correlation
measurements performed in the near and far fields. Entanglement in continuous variables (CV) is a rich research
subject, because several quantum information tasks can be
optimized using high dimensional Hilbert spaces \cite{bechmann00a,collins02}. SPDC is a
natural option for the experimental investigation of
transverse spatial entanglement, which can be present in other quantum
systems \cite{fedorov05,lamata07}.

So far, CV transverse entanglement detection has been based on
intensity correlation measurements performed in the near and in
the far field \cite{howell04,dangelo04}. An interesting
entanglement ``migration" effect was shown recently by Chan et. al
\cite{Chan07}, in which entanglement moves from the real to the
imaginary part of the two-photon wave function during propagation.
In order to be able to detect entanglement in this case, it would
be necessary to perform phase-sensitive measurements
\cite{peeters07}.

In this Letter, we show theoretically and experimentally that it
is always possible to detect entanglement by performing intensity
correlation measurements, even outside near and far field zones.
We demonstrate the connection between the variances of two
observables and the variances of these same observables rotated in
phase space. We encounter the conditions for which entanglement
detection is possible with intensity measurements, and others for
which it is impossible.  This connection allows one to circumvent
problems like the migration of entanglement \cite{Chan07} by
performing proper phase space rotations on the observables. Though
we consider the particular case of propagation of transverse
correlations of photon pairs, our results can be used to improve
detection of entanglement in other CV systems.

Our approach is based on the propagation of the signal and idler
fields using the formalism of the Fractional Fourier Transform
(FRFT), which is parameterized by the angle $\alpha$
\cite{ozaktas00}. The FRFT appears naturally in a number of
physical systems and describes rotation in phase space. In
particular, it is possible to completely describe the propagation
of a light field through the order $\alpha$ of the FRFT.  For
instance, the field at the source is given by a FRFT of order
$\alpha = 0$ and the usual Fourier transform, associated to
Fraunhoffer diffraction in the far field, is given by an FRFT of
order $\alpha = \pi/2$.  Free propagation can always be described
in terms of an FRFT operation up to a quadratic phase term, which
can be considered essentially unity in the near and far field
\footnote{In general, this phase term can be removed by
considering propagation to a spherical surface, and higher order
FRFT's can be defined by free propagation between sets of
spherical emitters and receivers \cite{pellat-finet94a,
pellat-finet94b}.}.  In Ref. \cite{howell04}, as is customary,
detection of entanglement was actually performed using lenses to
obtain the intensity correlations in the near field image ($\alpha
= \pi$) and far field ($\alpha = \pi/2$).  Likewise, any FRFT of
order $\alpha$ can be implemented perfectly with lenses
\cite{lohmann93,ozaktas00}.
\par
\begin{figure}
\centering
\includegraphics*[width=8.5cm]{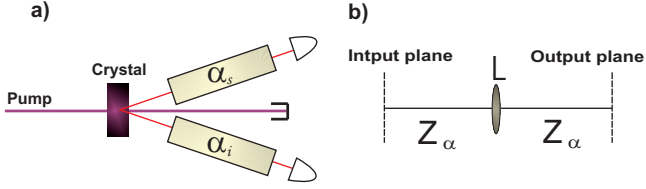}
\caption{\label{FIG:SimpleSetup2}(Color online) a) Experimental
Setup. Boxes on signal and idler paths represent the optical
systems used to perform FRFT of order $\alpha_s$ and $\alpha_i$. b) The optical lens system used to perform the FRFT \cite{lohmann93}. $L$ is a lens with focal
length ``$f$" and $z_{\alpha}=2f\sin^2(\alpha/2)$.}
\end{figure}

We consider the experimental arrangement sketched in Fig.
\ref{FIG:SimpleSetup2} a), where signal and idler photons from
SPDC are sent through FRFT systems of order $\alpha_s$ and $\alpha_i$, respectively.
\par
Following Ref. \cite{Chan07}, we write the two-photon wavefunction
at the source in position representation as
\begin{equation}\label{EQ:TwopartWaveFuncPOS}
    \bra{\rho_s,\rho_i}\Psi\rangle=
    \frac{1}{\left( {\pi}\sigma_-\sigma_+\right)^{1/2}}
    e^{-\frac{(\rho_i+\rho_s)^2}{4\sigma_+^2}}
   e^{-\frac{(\rho_i-\rho_s)^2}{4\sigma_-^2}},
    \end{equation}
where $\ket{\Psi}$ is the two photon quantum state produced by SPDC, $\rho_s$ and $\rho_i$
are appropriate dimensionless position coordinates in the source plane and $\sigma_{+}$ and $\sigma_{-}$ are independent functions of experimental parameters.
Once the state at the source is taken to be approximately gaussian, we only have to consider
one spatial dimension.  The two-photon wavefunction in wavevector representation is
\begin{equation}\label{EQ:TwopartWaveFuncMOM}
    \bra{q_s,q_i}\Psi\rangle=\sqrt{\frac{\sigma_+ \sigma_-}{\pi}}
    e^{-\frac{\sigma_+^2}{4}(q_i+q_s)^2}
    e^{-\frac{\sigma_-^2}{4}(q_i-q_s)^2},
\end{equation}
where $q_s$ and $q_i$ are transverse dimensionless wavevector components at the source plane.
\par
In order to detect entanglement of the state $\ket{\Psi}$, one must apply
a separability criteria \cite{duan00,mancini02}. For example,
defining the dimensionless operators $\hat{\rho}_{\pm}\equiv\hat{\rho}_i \pm \hat{\rho}_s$ and $\hat{q}_{\pm}\equiv\hat{q}_i \pm \hat{q}_s$ ($[\hat{\rho}_j,\hat{q}_k]=i\delta_{j,k}$ $j,k=s,i$)
the separability criteria of Duan, Giedke, Cirac and Zoller (DGCZ) \cite{duan00} establishes that if one
of the two inequalities $\langle
(\Delta\hat{\rho}_-)^2 \rangle_{\Psi} + \langle
(\Delta\hat{q}_+)^2 \rangle_{\Psi}\geq2$ or $\langle
(\Delta\hat{\rho}_+)^2 \rangle_{\Psi} + \langle
(\Delta\hat{q}_-)^2 \rangle_{\Psi}\geq2$ is violated, then the state
$\ket{\Psi}$ is non separable and therefore is entangled.

Using (\ref{EQ:TwopartWaveFuncPOS}) and (\ref{EQ:TwopartWaveFuncMOM}), we have
\begin{equation}
\label{EQ:Variances}
    \begin{array}{c}
      \langle \Delta(\hat{\rho}_+)^2 \rangle_{\Psi}= \sigma_+^2,\\
      \langle \Delta(\hat{\rho}_-)^2 \rangle_{\Psi}=\sigma_-^2,\\
      \langle \Delta(\hat{q}_+)^2 \rangle_{\Psi}= 1/\sigma_+^2,\\
      \langle \Delta(\hat{q}_-)^2 \rangle_{\Psi}= 1/\sigma_-^2,
    \end{array}
\end{equation}
and we obtain
\begin{equation}\label{EQ:DUANS}
    \langle (\Delta\hat{\rho}_-)^2 \rangle_{\Psi} +
\langle (\Delta\hat{q}_+)^2 \rangle_{\Psi}= \sigma_-^2 +
\frac{1}{\sigma_+^2}\;\;.
\end{equation}
The right hand side (RHS) of Eq. (\ref{EQ:DUANS}) can be smaller than 2
for small $\sigma_-$ and large $\sigma_+$.  In the
case of SPDC, this is readily achievable, as these two parameters
are independent and experimentally accessible.
\par
The variances in inequality \eqref{EQ:DUANS} refer to position and
momentum variables of the signal and idler fields in the source plane, which are related to the intensity distributions in the near and far field.
It is well known that the propagation of a light field
characterized by a FRFT is equivalent to a rotation of the
transverse variables in phase space \cite{ozaktas00}, given
that these variables are properly adimensionalized
\cite{pellat-finet94b}. The dimensionless operators transform as
\begin{equation}\label{EQ:RotatedOPER}
\begin{array}{c}
  \hat{\rho}_j \rightarrow \hat{\rho}_{\alpha_j}=
\cos\alpha_j\hat{\rho}_j +\sin\alpha_j\hat{q}_j \\
  \hat{q}_j
\rightarrow \hat{q}_{\alpha_j}= -\sin\alpha_j\hat{\rho}_j
+\cos\alpha_j\hat{q}_j ,\\
\end{array}
\end{equation}
where $j=s,i$. Therefore, it is  possible to write the DGCZ
inequality for rotated transverse variables of the fields,
$\hat{\rho}^{\prime}_- \equiv \hat{\rho}_{\alpha_s} -
\hat{\rho}_{\alpha_i}$ and $\hat{q}^{\prime}_+ \equiv
\hat{q}_{\alpha_s} + \hat{q}_{\alpha_i}$,
in terms of the variables $\hat{\rho}_-$
and $\hat{q}_+$ at the source \cite{tasca08b}:
\begin{eqnarray}\label{EQ:VariancesPRIME}
    &\langle (\Delta\hat{\rho}^{\prime }_-)^2 \rangle_{\Psi} + \langle
(\Delta\hat{q}^{\prime}_+)^2 \rangle_{\Psi}= \\ \nonumber
&\frac{1+ \cos(\alpha_i + \alpha_s)}{2} \left[
\langle(\Delta\hat{\rho}_-)^2 \rangle_{\Psi} + \langle
(\Delta\hat{q}_+)^2 \rangle_{\Psi} \right]\\ \nonumber & +
\frac{1- \cos(\alpha_i + \alpha_s)}{2} \left[
\langle(\Delta\hat{\rho}_+)^2 \rangle_{\Psi} + \langle
(\Delta\hat{q}_-)^2 \rangle_{\Psi} \right]\\ \nonumber & -
\frac{\sin(\alpha_i + \alpha_s)}{2}\left[ \langle\{ \hat{\rho}_+,
\hat{q}_+ \}\rangle - 2\langle \hat{\rho}_+\rangle \langle
\hat{q}_+\rangle  \right]_{\Psi} \\ \nonumber & +
\frac{\sin(\alpha_i + \alpha_s)}{2}\left[ \langle \{\hat{\rho}_-
,\hat{q}_-\} \rangle - 2\langle \hat{\rho}_-\rangle \langle
\hat{q}_-\rangle \right]_{\Psi}.
\end{eqnarray}
Eq. (\ref{EQ:VariancesPRIME}) shows that whenever $\alpha_i +
\alpha_s (\mbox{mod}\,2\pi) = 0\,$, the sum of variances for the
rotated variables coincide with the sum of variances for the
variables at the source. This shows that, for any propagation of
the signal field, characterized by $\alpha_s$, it is possible to
find a propagation of the idler field $\alpha_i$, so that
intensity correlation measurement will violate the DGCZ
inequality, in or out of the near and far field. We also note that
Eq. (\ref{EQ:VariancesPRIME}) does not depend on the state
(Eq.\ref{EQ:TwopartWaveFuncPOS}) and is applicable to any
bipartite continuous variable systems.

For states of the form (\ref{EQ:TwopartWaveFuncPOS})
the last two lines of the RHS of Eq. \eqref{EQ:VariancesPRIME} are zero.
Then considering an entangled state satisfying
$ \langle (\Delta\hat{\rho}_-)^2 \rangle_{\Psi} +
\langle (\Delta\hat{q}_+)^2 \rangle_{\Psi}= \sigma_-^2 +
1/\sigma_+^2\le 2$, a necessary condition to detect entanglement is
\begin{equation}
\cos(\alpha_i + \alpha_s) > \frac{S_1+S_2-4}{S1-S2}\geq 0,
\label{condition-no-violation}
\end{equation}
where we define $S_1\equiv \sigma_+^2+1/\sigma_-^2 $ and
$S_2\equiv \sigma_-^2+1/\sigma_+^2$.
We note that for $\cos(\alpha_i + \alpha_s)=0$, intensity correlation measurements never evidence entanglement, regardless of the state.
\par
We have experimentally tested these conditions, using pairs of twin
photons generated by SPDC in a 5mm
long lithium iodate crystal (LiIO$_3$) with a c.w. diode laser oscillating
at 405nm, as shown in FIG. \ref{FIG:SimpleSetup2} a).
Optical FRFT systems, such as the one shown in FIG. \ref{FIG:SimpleSetup2} b) were used in each of the down-converted fields.  This system, with $z_{\alpha}=2f\sin^2(\alpha/2)$, is able to implement a FRFT in the range $0\leq
\alpha\leq\pi$.  For $\alpha>\pi$ we use a series of FRFT systems,
respecting the additivity condition of maintaining $f^{\prime}=f \sin\alpha$ \cite{ozaktas00} the same for each.  To describe all FRFTs as rotations in the same phase space, we use dimensionless coordinates  $\rho=\sqrt{{k}/{f^{\prime}}}\,\bar{\rho}$ and
$q=\sqrt{{f^{\prime}}/{k}}\,\bar{q}$.  In our experimental setup,  $f^\prime=25/\sqrt{2}$\,cm.
Signal and idler photons were detected with single photon counting modules and
10nm FWHM bandwidth interference filters centered at 810nm. Horizontal slits ($100\mu$m)
were mounted on translation stages and scanned vertically in steps of $50\mu$m to register the detection position.   In all measurements, the ``+" (``-") correlations were measured in all cases by scanning the detectors with equal steps in the same (opposite) directions.
\par
First, we measured the $\rho_-$ and $q_+$ distributions at the source, using imaging ($\alpha_s=\alpha_i=\pi$) and Fourier transform ($\alpha_s=\alpha_i=\pi/2$) lens configurations \cite{howell04}.    The dimensionless variances were $\Delta^2(\rho_-)=0.93 \pm 0.01$ and
$\Delta^2(q_+)=0.073 \pm 0.004$.
Applying the DGCZ inequality  we obtain
\begin{equation}
\Delta({\rho}_-)^2
+ \Delta({q}_+)^2 =1.00 \pm 0.01 \le 2,
\label{eq:pivar}
\end{equation}
 indicating that the state is entangled.
\par
\begin{figure}
\centering
\includegraphics*[width=8.5cm]{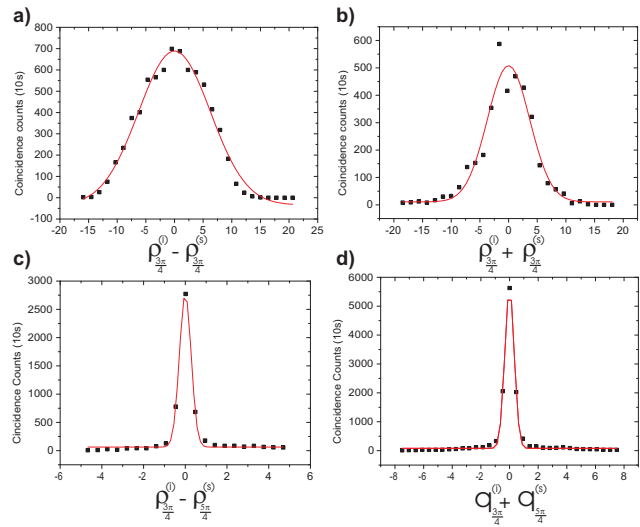}
\caption{\label{FIG:graficos} (Color online) Measured coincidence counts and gaussian curve fits. a)
$C(\rho_{\frac{3\pi}{4}}^{i}-\rho_{\frac{3\pi}{4}}^{s})$ b)
$C(\rho_{\frac{3\pi}{4}}^{i}+\rho_{\frac{3\pi}{4}}^{s})$ c)
$C(\rho_{\frac{3\pi}{4}}^{i}-\rho_{\frac{5\pi}{4}}^{s})$ d)
$C(q_{\frac{3\pi}{4}}^{i}+q_{\frac{5\pi}{4}}^{s})$}
\end{figure}

Next, we measured the intensity correlations for the signal and
idler fields at intermediate zones.    We chose FRFT orders $\alpha_s=\alpha_i=3\pi/4$, so that  $\cos(\alpha_s+\alpha_i)=0$, which does not satisfy the condition of Eq.(\ref{condition-no-violation}).
 The coincidence
counts  $C(\rho_{\frac{3\pi}{4}}^{i} - \rho_{\frac{3\pi}{4}}^{s})$ and
$C(\rho_{\frac{3\pi}{4}}^{i} + \rho_{\frac{3\pi}{4}}^{s})$ are
plotted in FIG. \ref{FIG:graficos}-a) and
(\ref{FIG:graficos}-b), respectively.
We obtain $\Delta^2(\rho_{\frac{3\pi}{4}}^{i} -
\rho_{\frac{3\pi}{4}}^{s})=13.6443>2$ and
$\Delta^2(\rho_{\frac{3\pi}{4}}^{i} +
\rho_{\frac{3\pi}{4}}^{s})=39.1473>2$, which clearly indicates that these
intensity correlations cannot be used to violate the DGCZ inequality.
We also tested an intermediate zone configuration following the condition
given by Eq. \eqref{condition-no-violation}.  We used three additive FRFT lens systems to
perform a $\alpha_s=\frac{5\pi}{4}$ order FRFT on the signal field, while maintaining the $\alpha_i=\frac{3\pi}{4}$ order FRFT on the idler field,
so that $\alpha_i +
\alpha_s = 2\pi$.  Coincidence counts
$C(\rho^{i}_{\frac{3\pi}{4}}-\rho^{s}_{\frac{5\pi}{4}})$ are
plotted in FIG. \ref{FIG:graficos}-c), and the dimensionless variance
is $\Delta^2(\rho_{\frac{3\pi}{4}}^{i}-{\rho_{\frac{5\pi}{4}}^{s}})=0.038 \pm 0.005$.
Coincidence counts
$C(q^{i}_{\frac{3\pi}{4}}+q^{s}_{\frac{5\pi}{4}})$, plotted in FIG. \ref{FIG:graficos}d), were measured
performing an inverse Fourier transform of the signal and idler
fields at the planes of FRFT of order $\frac{5\pi}{4}$ and
$\frac{3\pi}{4}$,
corresponding to FRFT of orders $\frac{3\pi}{4}$ and $\frac{\pi}{4}$, respectively.
The dimensionless variance is $\Delta^2(q^{i}_{\frac{3\pi}{4}}+q^{s}_{\frac{5\pi}{4}})=0.069 \pm 0.003$.
With our experimental data, we are now able to verify entanglement at intermediate zones:
\begin{equation}\label{EQ:RESULTS}
    \Delta^2(\rho_{\frac{3\pi}{4}}^{i}-{\rho_{\frac{5\pi}{4}}^{s}})
      + \Delta^2(q^{i}_{\frac{3\pi}{4}}+q^{s}_{\frac{5\pi}{4}})= 0.107 \pm 0.006 <2.
     \end{equation}
\par
 The experimental values obtained in Eqs. \eqref{eq:pivar} and
\eqref{EQ:RESULTS} are not equal as expected from Eq.
\eqref{EQ:VariancesPRIME}.
This discrepancy can be explained by the experimental imperfections.
It is difficult to characterize every source of
experimental error and their precise effect on the
measurement results. However, we notice that these imperfections
contribute by broadening the coincidence distributions.
In this respect, our measurement results are upper limits to the actual
variances.
To evaluate the expected variances, we
characterized the initial state (\ref{EQ:TwopartWaveFuncPOS}) at
the source by measuring the width $w$ of the intensity
distribution of the pump beam. The dimensionless variance at the
source is $\sigma_+^2=(4w^2){k/f^\prime}=47 \pm 2$. The
dimensionless variance $\sigma_-^2$ is given by
$\sigma_-^2=(k/f^\prime) 0.455 D/K=0.006$, where $D$ is the length
of the nonlinear crystal and $K$ is the pump beam wavenumber
\cite{Chan07}.  With these values, we predict a violation of the
DGCZ inequality:  $\langle (\Delta\hat{\rho}_-)^2 \rangle_{th} +
\langle (\Delta\hat{q}_+)^2 \rangle_{th} = 0.027 \pm 0.001 \le
2$, which is smaller than both experimental values
\eqref{eq:pivar} and \eqref{EQ:RESULTS}.
Therefore it is
clear that without the experimental imperfections, we should have observed
even stronger violations for both cases.
A possible imperfection is the error in lens positioning.
The effect of this type of imperfection can be estimated  by calculating the
propagation through the different lens systems in each field \cite{ozaktas00} and
including a 1\% error in z for all lens systems.  Taking the worst case scenario, we obtain the following predictions for each variance:
$\Delta^2(\rho_{\frac{3\pi}{4}}^{i}-{\rho_{\frac{5\pi}{4}}^{s}})_{th} = 0.09$ and
$\Delta^2(q^{i}_{\frac{3\pi}{4}}+q^{s}_{\frac{5\pi}{4}})_{th} = 0.04$, $\Delta^2(\rho_{-})_{th} = 0.84$ and
$\Delta^2(q_{+})_{th} = 0.03$.   These variances are much closer to the experimental  values.    Thus, considering small experimental imperfections, the theoretical prediction agrees with both Eqs. \eqref{eq:pivar} and \eqref{EQ:RESULTS}, as expected from  Eq.
\eqref{EQ:VariancesPRIME}.

\par
Let us now discuss the application of these results to a situation similar to that of Ref.\cite{Chan07}, in which it is shown that the transverse
intensity correlations decrease as the field propagates, and then are recovered again in the far-field.  For a certain propagation distance, the DGCZ or similar inequality will be satisfied, because the real part of the wavefunction becomes separable and
the entanglement is present only in the imaginary part.   An analysis similar to that
of Ref.\cite{Chan07} in terms of FRFT's yields a separability condition for the real part of the
wavefunction \eqref{EQ:TwopartWaveFuncPOS} given by
$\alpha_{sep}=\tan^{-1}(\sigma_+\sigma_-)$, where
$\alpha_s=\alpha_i=\alpha_{sep}$ is the order of the FRFT
implemented on both fields. Substituting this condition in Eq. \eqref{condition-no-violation} shows that the DGCZ inequality is not violated.
\par
\begin{figure}
\centering
\includegraphics*[width=6cm]{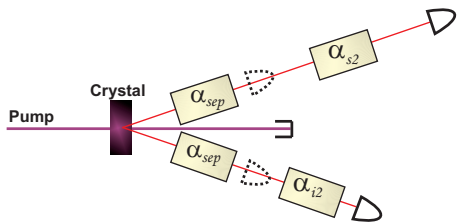}
\caption{\label{FIG:AlphaSEP} (Color online) Both signal and idler
fields propagate through lens systems that implement a FRFT of
order $\alpha_{sep}=\tan^{-1}(\sigma_+ \sigma_-)$. The dashed
detectors will observe no intensity correlation.
Additional FRFT systems are used so that the
total propagation for signal and idler are characterized by FRFT's
of orders $\alpha_{s} = \alpha_{sep}+\alpha_{s2}$ and $\alpha_{i}
= \alpha_{sep}+\alpha_{i2}$, respectively. Intensity correlation
is recovered for any $\alpha_{s2}$ and $\alpha_{i2}$ such that
$\alpha_i + \alpha_s (\mathrm{mod} 2\pi)= 0$.}
\end{figure}

To successfully detect entanglement in this case, it is necessary to adopt a scheme such as the one shown in Fig.
\ref{FIG:AlphaSEP}. Signal and idler fields propagate through optical
systems characterized by FRFT's of order $\alpha_{sep}$, so that the separability condition for the real
part of the wavefunction is reached.  At this point, intensity correlations alone will fail to register entanglement.  To detect entanglement, additional optical systems are
used to implement a second FRFT in the signal and
idler beams, with orders $\alpha_{s2}$ and $\alpha_{i2}$
respectively. The optical systems can be designed so that the
additivity of two consecutive FRFTs is preserved \cite{ozaktas00}.
The signal beam has had a total propagation
characterized by a FRFT of order $\alpha_{s} =
\alpha_{sep}+\alpha_{s2}$ and the idler $\alpha_{i} =
\alpha_{sep}+\alpha_{i2}$. In this case, applying the condition
$\alpha_i + \alpha_s (\mathrm{mod} 2\pi)= 0$, Eq. \eqref{EQ:VariancesPRIME} shows that the DGCZ inequality will be violated
in the same way  as it would be for the field in the source.

%
In conclusion, we have demonstrated theoretically and
experimentally, that it is possible to detect transverse
entanglement performing intensity correlation measurements, not
only in the near and far fields, but also at intermediate
propagation planes. This is achieved using optical systems that
implement Fractional Fourier Transforms (FRFTs) according to the condition
$\alpha_s+\alpha_i = 2n\pi$, where $\alpha_s$ and $\alpha_i$ are the
orders of the transforms on the signal and idler fields. We also show that entanglement
is never registered when $\alpha_i + \alpha_s (\mathrm{mod}\, \pi)= \pi/2$.  These results demonstrate that, given a signal field
propagation characterized by $\alpha_s$, one can always find an FRFT $\alpha_i$ which can be used to detect entanglement with intensity correlations alone.  Though our experiment was conducted using spatial entanglement of photons, our results are directly applicable to spatial entanglement in other systems\cite{fedorov05,lamata07}.  Since the Fractional Fourier Transform describes rotation in phase space, our results are applicable to a number of physical systems.

Financial support was provided by Brazilian agencies CNPq, PRONEX,
CAPES, FAPERJ, FUJB and the Milenium Institute for Quantum
Information.


\end{document}